\documentclass[pre,showpacs,superscriptaddress,floatfix]{revtex4}
\usepackage{epsfig,amsmath}
\begin{document}
\title{Charge and magnetic order in the 2D Falicov-Kimball
model with Hund coupling\\ at finite temperatures}
\author{Romuald Lema\'{n}ski}
\affiliation{Institute of Low Temperature and Structure Research, \\
Polish Academy of Sciences, Wroc\l aw, Poland}

\begin{abstract}
We perform a simplified analysis of finite temperature properties of the
generalized Falicov-Kimball model with Hund coupling. The model is studied
on the square lattice for fixed values of \emph{f-electron} and
\emph{d-electron} densities $\rho_f=1/2$, $\rho_d=1$.
Using an approxiamte scheme we estimated free energies of the following
three phases: the charge and spin ordered (SCO), the charge ordered (CO)
and the non-ordered (NO). Comparing the free energies we detected
phase transitions and found how the transition temperatures depend on 
the on-site interaction parameters. It appears, that the transition 
temperture between SCO and CO is much lower than between CO and NO,
but their maximum values are attained at values of coupling constants
very close to each other.

\end{abstract}
\maketitle

\section{Introduction}
Charge nad magnetic order are observed in many correlated electron systems
as, for example, in $R_{2-x}Sr_xNiO_4$, where $R=La,Nd$ \cite{RKajimoto}.
One of the models capable to describe such superstructures
is the Falicov-Kimball model, extended by the spin-dependent local interaction,
that reflects the first Hund's rule \cite{RLemanski}.

Until now the model has been studied only at zero temperature. 
Using the method of restricted phase diagrams or small cluster numerical
diagonalization it was shown how the charge distribution and spin arrangement 
change with parameters of the local interaction and densities of localized 
and itinerant electrons \cite{RLemanski,PFarkasovskyHCencarikova}. 
In particular, for $\rho_f=1/2$ and $\rho_d=1$ it was found that in 2D the ground state 
forms the checkerboard charge pattern consistent with the simplest antiferromagnetic 
order \cite{RLemanski}.

Here we investigate behaviour of the system at finite temperatures.
Our purpose is to determine the way of transformation of the ordered phase
into disordered one when temperature increases. The transformation appears 
to split into two phase transitions: at $T_{MO}$ between the low temperature phase, 
where both charge and spin are ordered (SCO) and the intermediate phase, with only 
charge but not spin ordered (CO), and then at $T_{CO}$ between CO and the high 
temperature phase, that is disordered with respect to both the spin and charge (NO).
A visualization of arrangements of the localized electrons in the three phases 
is shown in Fig. 1.
\begin{figure}[htb]
\begin{center}
    \epsfxsize=14cm
    \epsffile{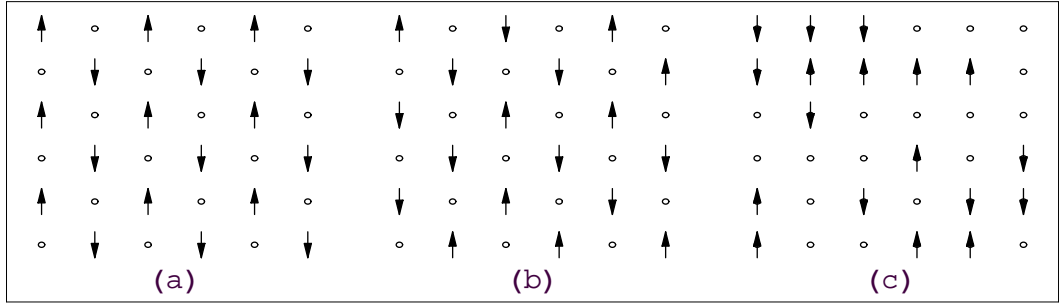}
\caption{Typical arrangements of the localized electrons (snapshots) in three different
phases: a) the spin and charge ordered (SCO), b) the charge-ordered (CO)
and c) the non-ordered (NO).}
\label{fig1}
\end{center}
\end{figure}

The model Hamiltonian is
\begin{eqnarray}
\label{ham}
H= & t\sum\limits _{\left\langle i,j \right\rangle}\sum\limits _{\sigma=\uparrow ,\downarrow}
d^+_{i,\sigma} d_{j,\sigma}
+U \sum\limits_{i}\sum\limits_{\sigma ,\eta =\uparrow,\downarrow}
n^d_{i,\sigma} n^f_{i,\eta} \nonumber \\
 & -J \sum\limits _{i}(n^d_{i,\uparrow}-n^d_{i,\downarrow} ) (n^f_{i,\uparrow}-n^f_{i,\downarrow} ),
\end{eqnarray}
where $<i,j>$ means the nearest neighbor lattice sites $i$ and $j$,
$\sigma$ and $\eta$ are spin indices, $d_{i,\sigma}$  ($d^+_{i,\sigma}$) is an annihilation
(creation) operator, and $n^d_{i,\sigma}$ ($n^f_{i,\sigma}$) is
an occupation number of itinerant(localized) electrons.
The on-site interaction between localized and itinerant
electrons is represented by two coupling constants: $U$, which is
spin-independent Coulomb-type and $J$, which is spin-dependent and reflects
the Hund's rule force. The hopping amplitude $t$ we set
equal to one, so we measure all energies in units of $t$.

Double occupancy of the localized electrons is forbidden,
implying the on-site Coulomb repulsion $U_{ff}$ between two
\emph{f-electrons} is infinite.
Consequently, at a given site the \emph{f-electron} occupancy is assumed
to be $n_f=n_{f,\uparrow} + n_{f,\downarrow} \leq 1$ and the
$d-electron$ occupancy to be $n_d=n_{d,\uparrow} + n_{d,\downarrow} \leq 2$.
So there are 3 states per site allowed for the \emph{f-electrons}
($n_f=0$; $n_{f,\uparrow}=1$ and $n_{f,\downarrow}=0$;
$n_{f,\uparrow}=0$ and $n_{f,\downarrow}=1$)
and 4 states per site allowed for the \emph{d-electrons}
($n_d=0$; $n_{d,\uparrow}=1$ and $n_{d,\downarrow}=0$;
$n_{d,\uparrow}=0$ and $n_{d,\downarrow}=1$; $n_d=2$).

All single-ion interactions included in (\ref{ham}) preserve states
of the localized electrons, i.e. the itinerant electrons traveling through
the lattice change neither occupation numbers nor spins of the localized
ones. In other words, $[H,f^+_{i\eta }f_{i\eta }]=0$ for all $i$ and
$\eta $, so the local occupation number is unchanged.

The localized electrons play the role of an external, charge and spin
dependent potential for the itinerant electrons. This external potential
is "adjusted" by annealing, so the total energy of the system
attains its minimum. In other words, there is a feedback between the subsystems
of localized and itinerant electrons, and this is the feedback
that is responsible for the long-range ordered  arrangements of the localized
ones, and consequently for the formation of various charge and/or spin
distributions in low temperatures.

\section{Method of calculation}
In this paper we consider only the case of $\rho_f=1/2$ and $\rho_d=1$.
Then, for the system composed of $N$ sites there are $\frac{N!}{[(N/2)!]^2}$ 
possible charge distributions and, for each of them, there exist $2^{N/2}$ 
spin configurations.
So the entropy (per site) resulting from various possible distributions
of \emph{f-electrons} is a sum of the charge part equal to 
$\frac{1}{N}Log\frac{N!}{[(N/2)!]^2}$,
and the magnetic part equal to $\frac{1}{2}Log2$.
In the limit of $N\rightarrow\infty$ the charge part of entropy
tends to $Log2$.

For any fixed configuration of localized electrons $C$ one can calculate all 
thermodynamic characteristics of the \emph{d-electron} subsystem from standard 
formulas (see for example \cite{MMaskaKCzajka}). 
In particular, the Gibbs potential per site $g_C$ is given by the expression
\begin{equation}
 g_C(\mu)=-\frac{1}{N\beta }\sum\limits _{n}Log\left(  1+e^{-\beta [E_n(C)-\mu]}\right) 
\end{equation}
and the free energy $f_C$ by
\begin{equation}
 f_C(\rho_d)=g_C(\mu(\rho_d))+\mu(\rho_d)\rho _d
\end{equation}
with the following condition fulfilled (in our case $\rho _d=1$):
\begin{equation}
 \rho_d=\frac{1}{N}\sum\limits _{n}\frac{1}{1+e^{\beta [E_n(C)-\mu]}}
\end{equation}
Then, the partition function per site $z$ is obtained from the summation over all
configurations $C$ of \emph{f-electrons}
\begin{equation}
z=\frac{1}{N}\sum\limits _{C}e^{-\beta f_CN}
\end{equation}
and the free energy $f$ is equal to
\begin{equation}
 f=-\frac{1}{N\beta}Log(\sum\limits_{C}e^{-\beta f_CN}).
\label{totfree}
\end{equation}
Since we are not able to determine $f$ exactly in the whole temperature range,
we replace it by three different approximate functions,
relevant for three different phases: CSO, CO and NO.
In order to do this we first divide the sum in (\ref{totfree}) as follows 
\begin{equation}
\sum\limits _C e^{-\beta N f_C}=e^{-\beta N f_{C_{SCO}}}+\sum\limits _{C\in C_{ch}^*} e^{-\beta N f_C}+\sum\limits _{C\in C^*} e^{-\beta N f_C}
\label{mainsum},
\end{equation}
where $f_{C_{SCO}}$ is the free energy of the SCO phase, $C_{ch}^*$ denotes 
the set of checkerboard configurations $C_{ch}$ (with respect to charge) 
of the \emph{f-electrons} with all possible spin arrangements but SCO
and $C^*$ denotes all C configurations but $C_{ch}$.
\begin{figure}[htb]
\begin{center}
    \epsfxsize=8cm
    \epsffile{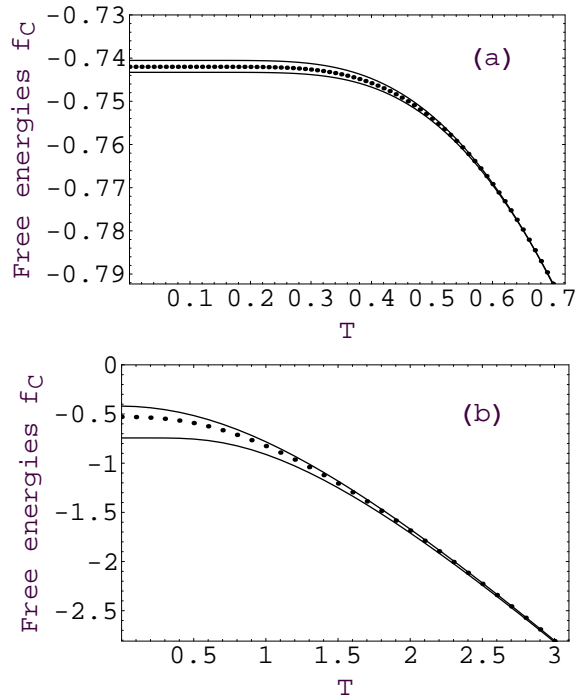}
\caption{Free energies $f_C$ (calculated per site) of phases with fixed
configurations of the \emph{f-electrons} (the continous lines) corresponding
to the highes and the lowest value functions taken from a choosen set
of configurations: a) $C\in C_{ch}$ b) all $C$. The points mark approximative average free
energy function taken in the calculations: a) $ \langle f_C \rangle_{C \in C_{ch}}$,
b) $ \langle f_C \rangle_{all C}$.}
\label{fig2}
\end{center}
\end{figure}

With an increase of temperature the subsequent terms in (\ref{mainsum})
contribute to $f$: for the lowest temperatures only the first term, 
then, for intermediate temperatures two first terms and, finally, for
higher temperatures all three terms.
So for temperatures very close to zero the main contribution to $f$ comes
from the free energy $f_{C_{CSO}}$, attributed to the ground state
configuration CSO of the \emph{f-electrons}
\begin{equation}
f \simeq -\frac{1}{\beta}Log(e^{-\beta f_{C_{CSO}}})=f_{C_{CSO}}.
\end{equation}
For higher temperatures we calculate mean free energies by averaging
them over appriopriate sets of \emph{f-electron} configurations.
In order to find the mean free energies we take into account a representative class
of configurations $C$ of the \emph{f-electrons} only, composed of all
periodic phases with periods not exceeded four lattice sites. 
Then our trial set contains 13 non-equivalent configurations.
Three of them belong to the $C_{ch}$ class. For all
these configurations we calculated exact free energies per site
in the limit of large N by solving the eigenvalue problem
and finding the eigenvalues $E_n(C)$. This required us to diagonalize
up to $4\times 4$ matrices and result in analytical formulae for at most
4 different energy bands, separately for spin-up and spin-down \emph{d-electrons}
(for more details see Refs. \cite{RLemanski,GWatsonRLemanski,LemanskiFreericksBanach}).

For intermediate temperatures we tak an average $f_C$ over the family $C_{ch}$ only. 
Their free energies $f_C$ are differrent, but the difference is quite small,
even for $T=0$, and tends to zero with temperature (see Fig. 2).

The whole contribution to $f$ coming from this family could be expressed 
by a mean value $ \langle f_C \rangle_{C \in C_{ch}}$ in the following way.
\begin{equation}
f \cong f_{CO}=-\frac{1}{N\beta}Log(2^{N/2}e^{-\beta N\langle f_C \rangle _{C \in C_{ch}}})=-\frac{1}{2\beta}Log2+ \langle f_C \rangle _{C \in C_{ch}}
\label{fco}
\end{equation}
We get the appropriate mean free energy $ \langle f_C \rangle_{C \in C_{ch}}$ using 
the arithmetical average of the relevant free energies $f_C$.
The averaging procedure we used is equivalent to taking the geometrical mean value
of relevant ingradients of the partition function. Indeed, we get (\ref{fco}) if we replace
each element of the sum $\sum\limits _{C\in C_{ch}} e^{-\beta N f_C}$
by $M_{C_{ch}}\cdot(\prod \limits _{C\in C_{ch}}e^{-\beta N f_C})^{1/M_{C_{ch}}}$,
where $M_{C_{ch}}=2^{N/2}$ is the number of configurations in  $C_{ch}$ 

The same procedure we use when calculating an approxiamtive mean free energy
$f_{NO}$ of the non-ordered phase. However, then we take an arithmetric average 
of free energies of all configurations from the choosen set, and take into account
the fact, that in the large N limit there are 
$2^{N/2}\frac{N!}{[(\frac{N}{2})!]^2}\approx 2^{N/2}2^N$ 
such configurations. So we have
\begin{equation}
 f \cong f_{NO}=-\frac{1}{2\beta}Log2-\frac{1}{\beta}Log2+\langle f_C\rangle_{allC}.
\end{equation}
We tested the averaging procedure for the spinless Falicov-Kimball model
and found, that the transition temperatures we got are nearly the same
as those obtained by the Monte Carlo simulations in the whole range
of the interaction parameter $U$ \cite{MMaskaKCzajka}.

\section{Results and conclusions}
The results of our calculations are displayed in Fig. 3. In the upper panel a)
it is shown the magnetic order-disorder transition temperature $T_{MO}$
between SCO and CO phases and in the lower panel b) the charge order-disorder
transition temperature $T_{CO}$ between the CO and NO phases.
Both $T_{MO}$ and $T_{CO}$ were calculated for $J=0.2U$ as functions of the
rescaled on-site interation parameter $\frac{U}{U+1}$.
\begin{figure}[htb]
\begin{center}
    \epsfxsize=10cm
    \epsffile{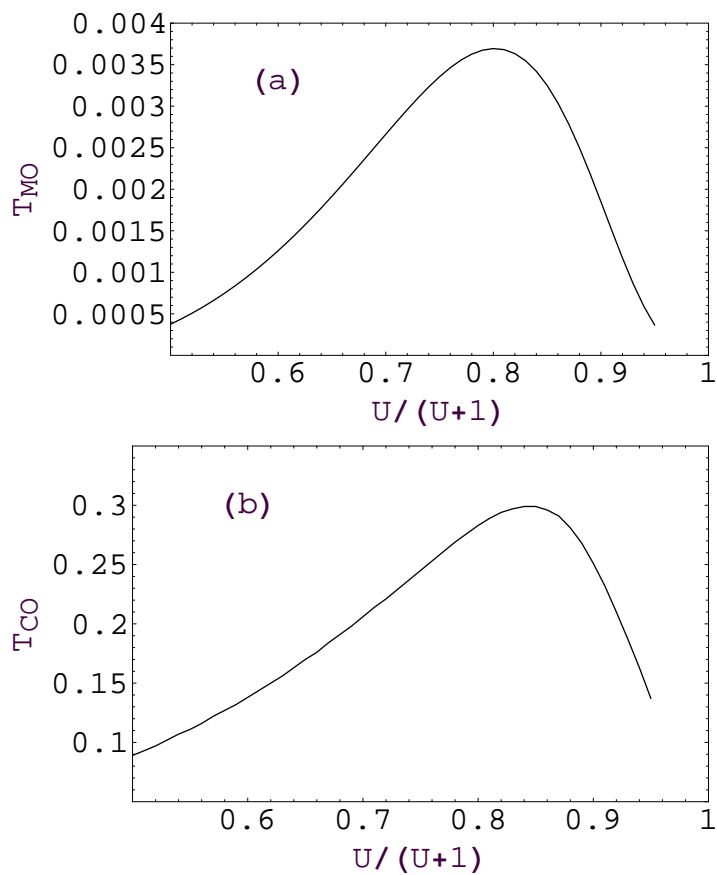}
    \caption{The temperatures of order-disorder phase transitions:
a) magnetic $T_{MO}$ and b) charge $T_{CO}$, as a function of the on-site
interaction parameter (in the both cases $J=0.2U$).}
\label{fig3}
\end{center}
\end{figure}

Since we used rather crude approximative scheme of calculations of the
free energies, the obtained values of the transition temperatures
could not be very precise. However, we expect that the quantitative picture
is correct. In particular, two main conclusions seem to be relevant.
First, the maxima of the both transition temperatures are attained
at very close to each other values of the parameter $\frac{U}{U+1}\approx0.8\div0.85$.
It corresponds to $U\approx4\div5.67$ (and $J\approx0.8\div1.13$, respectively).

Another important message resultant from the calculations is the fact,
that $T_{MO}$ appears to be almost two orders of magnitude lower than $T_{CO}$,
even though the Hund coupling constant $J$ is assumed to be only
5 times smaller than $U$. The possible cause of this rather unexpected
outcome is the lack of direct interaction between the spin-up 
and spin-down itinerant electrons (the Hubbard term) in the model. We anticipate that
the inclusion of this type of interaction would enhance the magnetic 
couplings in the system, resulting in an increase of the magnetic 
transition temperature.

\acknowledgements
We acknowledge support from the Polish Ministry of Sciences and Higher Education
under the grant no. 1 P03B 031 27.

\end{document}